\documentclass[10pt]{asme2ej}

\usepackage{epsfig} 
\usepackage{amsmath}
\usepackage{amssymb}
\usepackage{mathptmx}
\usepackage{graphicx}
\usepackage{amssymb}
\usepackage[mathscr]{euscript}
\usepackage{times} 
\usepackage{lipsum} 
\usepackage{epstopdf}
\usepackage{pstricks}
\usepackage{ulem}
\usepackage{verbatim}
\usepackage{fancyhdr}
\renewcommand{\headrulewidth}{0pt}
\renewcommand{\footrulewidth}{0.4pt}

\title{Clinical Facts Along With a Feedback Control Perspective Suggest That Increased Response Time Might be the Cause of Parkinsonian Rest Tremor}

\author{Vrutangkumar V. Shah
    \affiliation{
	Doctoral Student\\
	SysIDEA Lab\\
	Mechanical Engineering\\
	Indian Institute of Technology Gandhinagar\\
	INDIA-382355\\
    Email: shah\_vrutangkumar@iitgn.ac.in
    }	
}

\author{Sachin Goyal
    \affiliation{ Assistant Professor\\
    SysIDEA Lab, IIT-GN\\
	Department of Mechanical Engineering\\
	University of California\\
	Merced, CA 95343\\
        Email: sachin.goyal@ucmerced.edu
    }
}

\author{Harish J. Palanthandalam-Madapusi
    \affiliation{
	Assistant Professor\\
        SysIDEA Lab\\
	Mechanical Engineering\\
	Indian Institute of Technology Gandhinagar\\
	INDIA-382355\\
	Email: harish@iitgn.ac.in
    }
}

\begin{document}

\maketitle    


\begin{abstract}
{\it Parkinson's disease (PD) is a neurodegenerative disorder characterized by increased response times leading to a variety of biomechanical symptoms such as tremors, stooping and gait instability. Although the deterioration in biomechanical control can intuitively be related to sluggish response times, how the delay leads to such biomechanical symptoms as tremor is not yet understood. Only recently has it been explained from the perspective of feedback control theory that delay beyond a  threshold can be the cause of Parkinsonian tremor \cite{11}. This paper correlates several observations from this perspective to clinical facts and reinforces them with simple numerical and experimental examples. This work provides a framework towards developing a deeper conceptual understanding of the mechanism behind PD symptoms. Furthermore, it lays a foundation for developing tools for diagnosis and progress tracking of the disease by identifying some key trends.
}
\end{abstract}

\section{Introduction}

Patients suffering from Parkinson's disease (PD) experience a variety of biomechanical symptoms including tremors, stooping, rigidity, and gait instability \cite{9}. Since the discovery of this disease in 1817 \cite{3}, the connections between these apparently unrelated biomechanical symptoms has puzzled researchers and have led to a range of hypotheses and conjectures about the source of these symptoms \cite{34, 35} and it is not clear whether there is a single underlying explanation for these symptoms \cite{49, 50}. PD is a neuro-degenerative disorder and is also characterized by a permanent increase in response time in both voluntary and involuntary motor responses. The impairment of response time in Parkinson's disease was first noted in \cite{2}, stating, ``Recent measurements with special apparatus for muscular response to a single visual stimulus have given the figures of 0.24 seconds for normal individuals and 0.36 seconds for the subjects with paralysis agitans". Several detailed studies  since then have come up with similar conclusions \cite{13, 14, 15, 17}. This has added another dimension to the mystery surrounding the symptoms.

Although the deterioration in biomechanical control can intuitively be related to sluggish response times, how the increase in response time leads to such biomechanical symptoms as tremors is not yet understood \cite{6}. In fact it is argued that Parkinsonian tremor may have a pathophysiology different from most other symptoms of the disease \cite{34, 35}. 
Neuroprosthetic therapies such as deep brain stimulation \cite{37} suppress Parkinsonian tremor, however, the fundamental mechanism behind these therapies is also unresolved \cite{38}.

On another front, empirical mathematical models are proposed that can simulate Parkinsonian tremor, for instance, a limit-cycle-exhibiting system such as the Van-Der-Pol oscillator can be fit to experimentally measured data \cite{10}. But such an approach lacks physical underpinnings and does not provide any insights into some of the other key features of Parkinsonian tremor. For example, why a PD patient trying to keep still would exhibit tremors (referred to as rest tremors) \cite{9}, whereas these tremors would often disappear during voluntary motion \cite{34}, is not explained by such models. Recently, arguments based on a control-system analogy were used to support the hypothesis that Parkinsonian tremor may indeed be a limit cycle oscillation \cite{11}, and established a direct logical connection between increased response time and limit-cycle behaviour of Parkinsonian tremor. Since then, similar connections between increased time delays and limit cycle oscillations in human biomechanics (although not necessarily in the context of Parkinson's disease) have also been drawn \cite{31, 45, 46}. 

\begin{figure}[h]
	\begin{center}
		\includegraphics[scale=0.2]{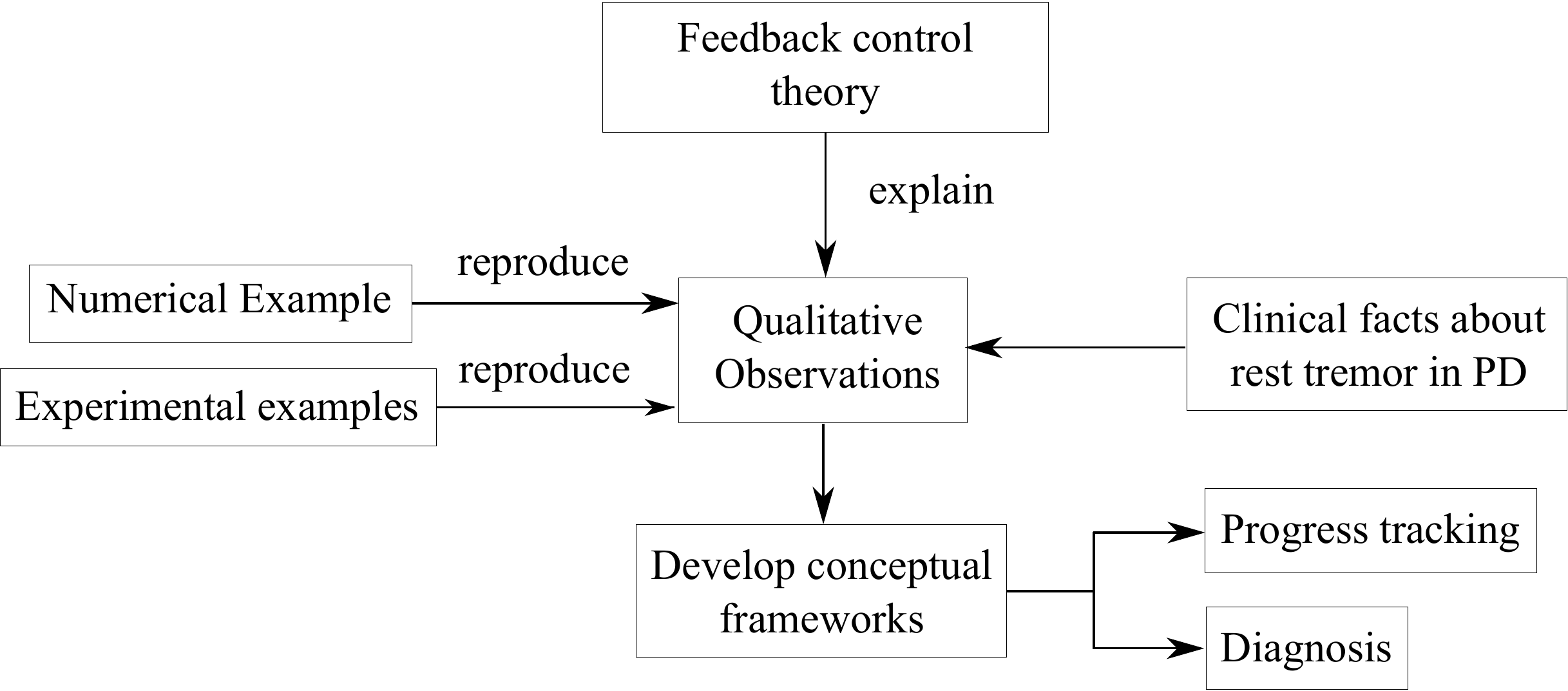}
	\end{center}
	\caption{The contributions of this work.\label{organization}}
\end{figure}

In this paper, we exploit this link between increased time delays (observed as an increase in response times) and Parkinsonian tremors to address two specific objectives (see Fig. \ref{organization}). First, we wish to draw qualitative observations based on this hypothesis that are supported by clinical facts, explained using feedback control theory, and reproduced by simple numerical and experimental examples. Second, based on this hypothesis, we explore possibilities for using biomechanical analysis of tremor data for diagnosis  and progress tracking of the disease.
The current work therefore builds a framework towards developing a deeper conceptual understanding of the mechanism behind Parkinsonian tremor. Further, we demonstrate how, in-principle, the insights in this paper can be used to develop a simple pocket device or smartphone application for progress tracking and diagnosis of the disease \cite{44}. For exploring the possibilities for diagnosis and progress tracking in further details, one can use a combination of patient tests and simulations with simple, yet reasonably realistic models of human posture, motor control, or gait \cite{33, 47} as the next steps.

The paper is organized as follows. Section \ref{representation} introduces motor control of human body in feedback control framework that will help us in subsequent analysis. Section \ref{Model} explores how some well-known clinical facts about Parkinsonian tremors can be explained using feedback control arguments and validated with simple numerical and experimental examples. Section \ref{diagnosis} explores the possibilities for diagnosis and progress tracking of the disease based on the results and then we
close with conclusion in Sec. \ref{conclusion}.

\section{Feedback Control Representation}
\label{representation}

\begin{figure}
\begin{center}
\includegraphics[scale=0.42]{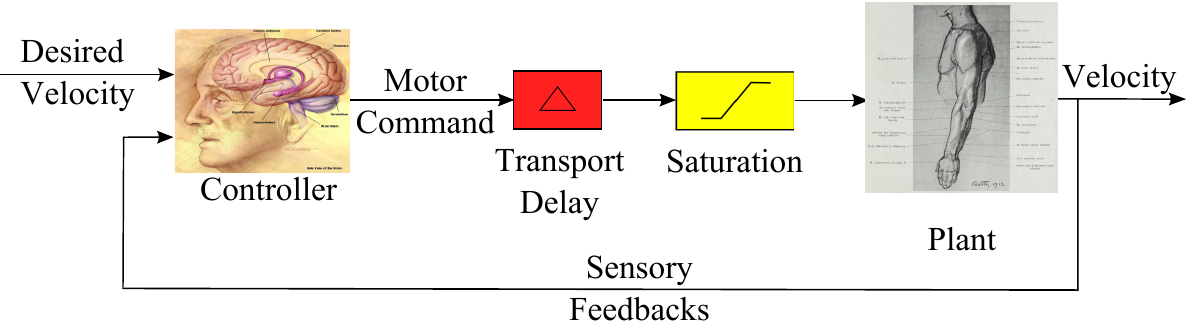}
\end{center}
\caption{ The closed-loop feedback system representing motor control of human body.\label{Block_Diagram}}
\end{figure}
For our analysis, we adopt the feedback control system framework presented in \cite{11} as shown in Fig. \ref{Block_Diagram}. This is a simple block-diagramatic representation of the motor control using sensory feedback in humans. In this framework, the mathematical model that governs the dynamics of any body part (e.g. hand) in the absence of any neural control is what is referred to as a plant in control-system notation. The feedback path represents all sensory feedbacks including visual feedback, tactile feedback, proprioceptive feedback, etc. Afferent nerves carry these feedback information to the controller (brain).  The controller represents the neurosystem's logic that continuously compares the kinematic variables from sensory feedbacks (e.g. actual velocity) with the desired kinematic variables (e.g. desired velocity) to determine motor command. The information of the particular command is carried by the efferent nerves and are then implemented on the plant via muscle actions to get the desired response.  We refer to this closed feedback loop consisting of motor actions and sensory feedbacks as the sensorimotor loop.
We expect that there will be various time delays in the sensorimotor loop such as delays due to nerve conduction times, information processing time, etc. For simplicity, we lump all sensorimotor loop delays (delays in various portions of sensorimotor loop) into one €transport delay in the closed-loop feedback system. Given that in PD we see larger response time compared to healthy individuals, it is reasonable to assume that this lumped delay is larger in patients with PD.  We discuss in Appendix A how under a linearity assumption the specific locations of the various delays do not affect the overall observations derived from the analysis that follows. Finally, the physiological limit of the transmission of neural control actions \cite{12} is represented as a saturation function that bounds the control input to the plant.
\section{Explaining Clinical Facts Using Feedback Control Representation}
\label{Model}

With the framework described in Sec. \ref{representation}, we explore how some well-known clinical facts about Parkinsonian tremor can be explained using feedback control arguments. Three examples that are explained in detail in Appendix B are chosen for this analysis; a numerical example and two table-top motion control experimental examples. The intention here is to choose three simple yet very different examples within the framework described in Sec. \ref{representation}, to demonstrate the generality of the insights and observations that follow. The insights and trends from these examples are subsequently leveraged to conceptualize strategies for diagnosis and progress tracking. However, it is to be noted that more realistic models of human motor control \cite{33,47} along with patients studies would be needed subsequently to further develop and fine-tune these diagnosis and progress tracking strategies. \\
\indent The numerical example is that of simple pendulum in MATLAB Simulink and is modeled as a second-order stable linear system with a proportional-integral-derivative controller. The first experimental example is a servo motor position control bench-top experiment, which is a first-order, linear, stable system, with a proportional-derivative controller. The second experimental example is a rotary inverted pendulum, which is a fourth-order, non-linear, unstable system, with an LQR controller. Although all three examples are different, they follow the same feedback control framework and all three have saturation and delay, the two crucial features to explain the possible mechanism of Parkinsonian tremor. Thus, together they serve to qualitatively explain the clinical facts by highlighting how the feedback control perspective can explain the facts regardless of variations between individuals. We describe results for the Pendulum numerical example and the Servo motor experimental example in more detail and show some representative results for the {Rotary inverted pendulum experimental example.

 The subsections that follow describe four clinical facts and the supporting analysis. \\

\subsection{ Parkinsonian tremor is primarily a rest tremor \cite{9}. \label{rest_tremor}}

\indent  Rest tremor is shaking of body parts, most commonly the hands or fingers, when these body parts are in rest position. By contrast, action tremor occurs during any type of intentional movement of an affected body part. In this context, all three of our simple examples revealed that a delay beyond a certain threshold triggers oscillatory behaviour for zero intended velocity (feedback control trying to bring the system to rest). For example,  Fig. \ref{sim1} shows that the Pendulum numerical example exhibits no oscillations in the steady-state response with the delay values of 0.05 s and 0.1 s (both responses are visually indistinguishable in Fig. \ref{sim1}). However, as the delay increases from 0.1 s to  0.15 s, the response settles at a steady-state oscillation, implying that the threshold delay is somewhere between 0.1 s and 0.15 s. Likewise, for the Servo motor experimental example, Fig. \ref{ex1} indicates that the threshold delay is somewhere between 0.03 s and 0.04 s.
These observations can be readily explained from feedback control theory in the following way. It is well known that any stable feedback control loop with loop gain more than one becomes unstable if a large enough delay is introduced in the loop \cite{18}. Further, due to the presence of the saturation in the loop, the instability is not able to drive the  amplitude of oscillations to infinity, but rather a bounded oscillatory behaviour is approached \cite{18}. In the phase space, an increased delay (beyond a certain threshold) renders the origin  (the equilibrium corresponding to the rest condition) unstable, and even with the smallest disturbance or initial condition, the response is driven to a stable limit-cycle oscillation due to the saturation in the loop. Figure \ref{limitcycle} (Pendulum numerical example), Fig. \ref{limitcycle1} (Servo motor experimental example) and Fig. \ref{rotary} (Rotary inverted pendulum experimental example) confirm this explanation by showing that all the trajectories starting from different initial conditions in the phase space plot of angular velocity versus angular acceleration converge to a stable limit cycle. Hence, these examples explain, in relatively simple terms, why Parkinsonian tremor is a rest tremor and occurs when the patient tries to keep still.

\begin{figure}
\begin{center}
\includegraphics[scale=0.25]{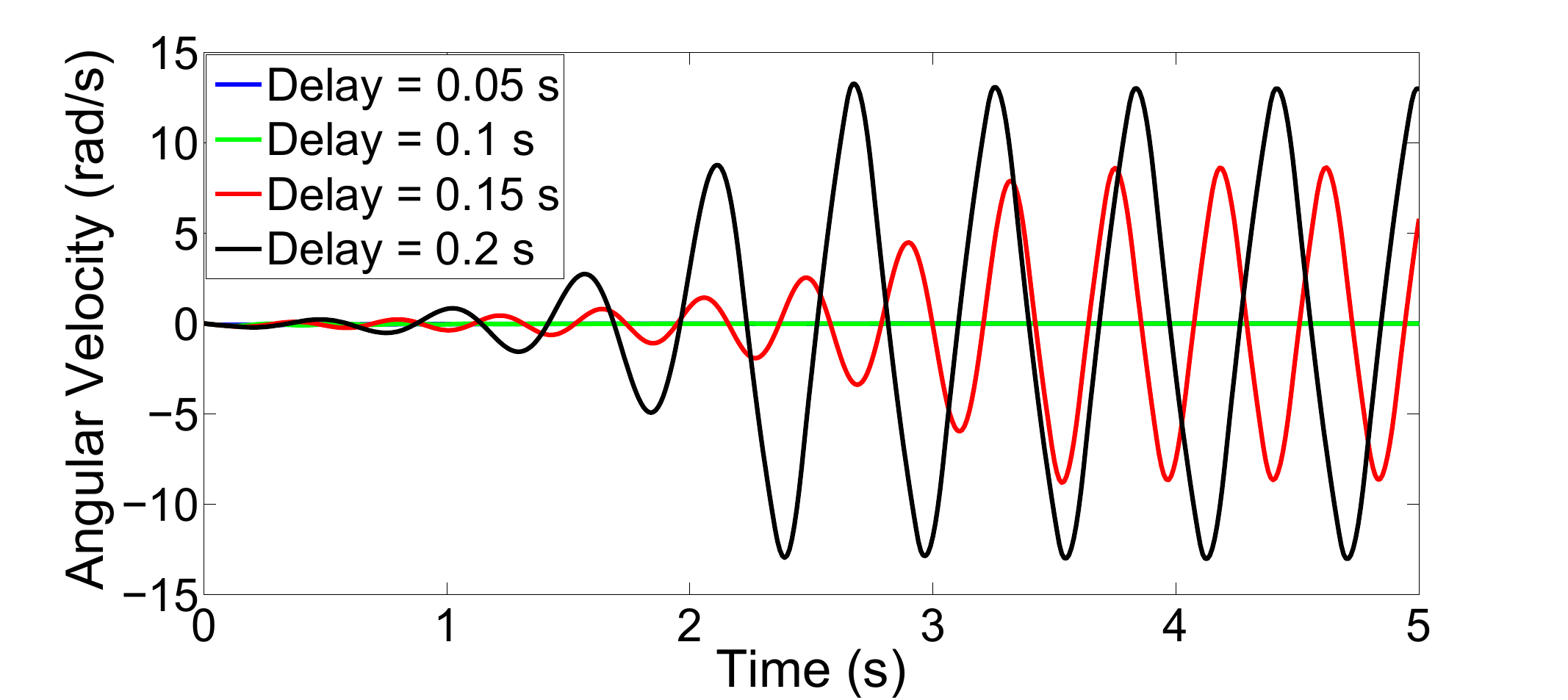}
\end{center}
\caption{Angular velocity for various time delays with zero intended velocity, initial angular position of $0.1$ $rad$ and saturation limits as $-100$ to $100$ $N$-$m$ (Pendulum numerical example). \label{sim1}}
\end{figure} 
\begin{figure}
\begin{center}
\includegraphics[scale=0.25]{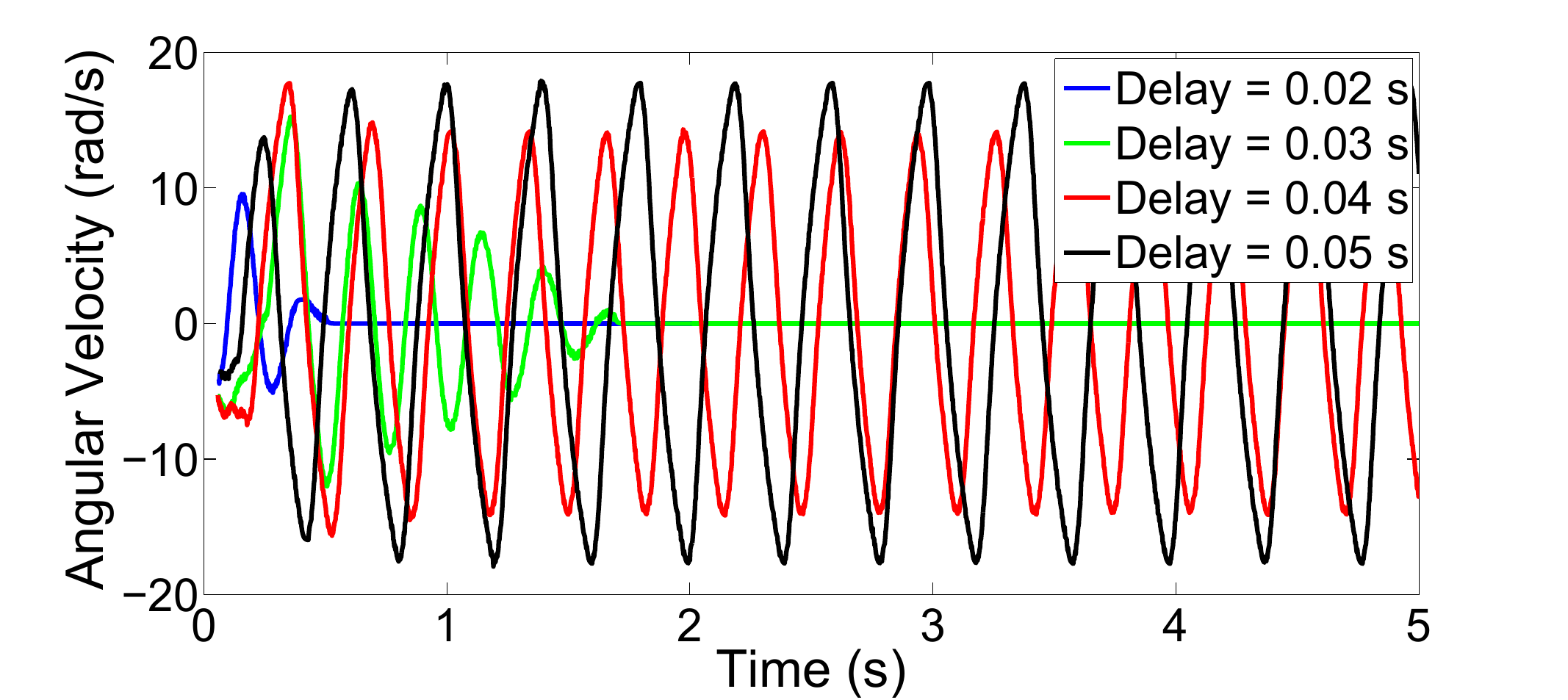}
\end{center}
\caption{Angular velocity for various time delays with zero intended velocity, various initial angular velocity ($-4.11$ $rad/s$ for delay = $0.02$ $s$, $-3.52$ $rad/s$ for delay = $0.03$ $s$, $-5.35$ $rad/s$ for delay = $0.04$ $s$, $-5.27$ $rad/s$ for delay = $0.05$ $s$) and saturation limits as $-1$ to $1$ $N$-$m$ (Servo motor experimental example). \label{ex1}}
\end{figure} 
\begin{figure}
\begin{center}
\includegraphics[scale=0.25]{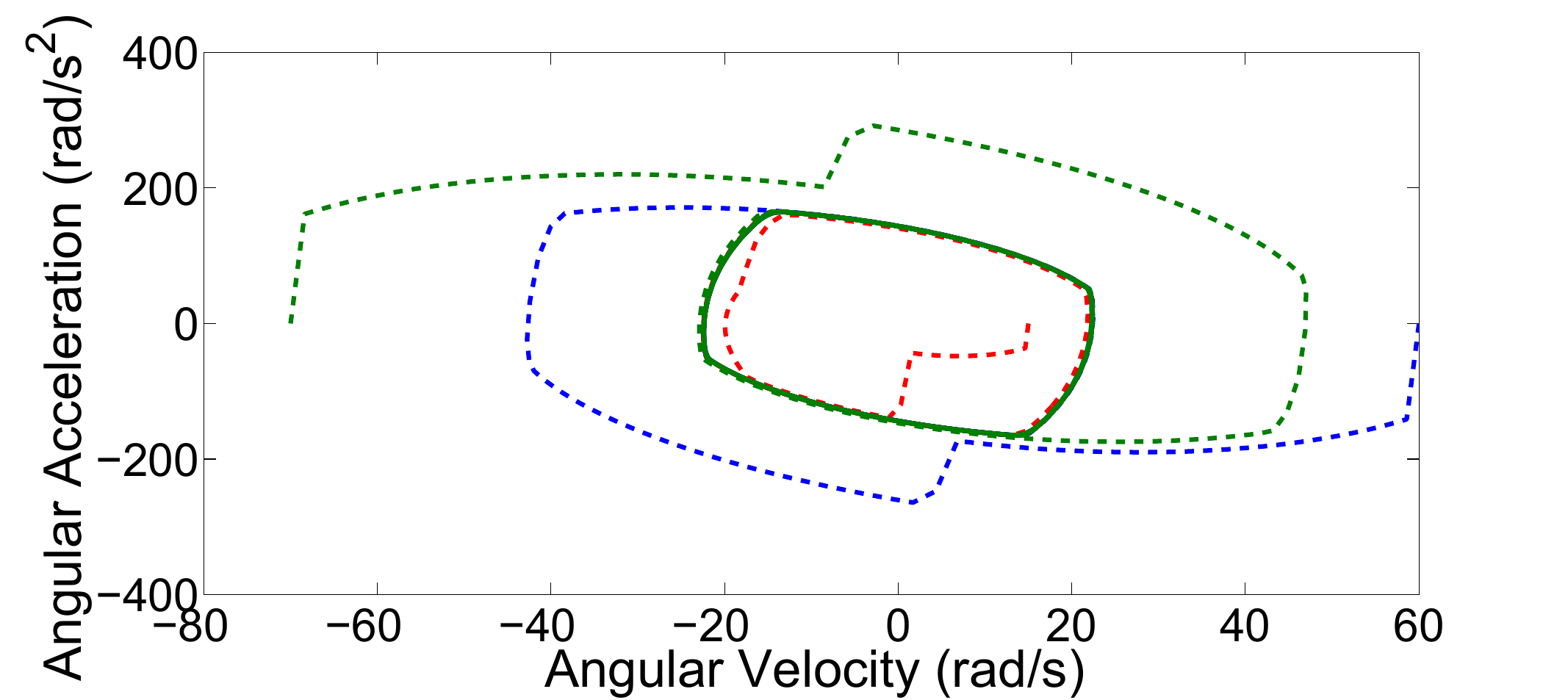}
\end{center}
\caption{Trajectories starting from various initial conditions converging to a limit cycle in phase space for delay $0.3$ $s$ and saturation limits $-100$ to $100$ $N$-$m$ (Pendulum numerical example). \label{limitcycle}}
\end{figure} 
\begin{figure}
\begin{center}
\includegraphics[scale=0.25]{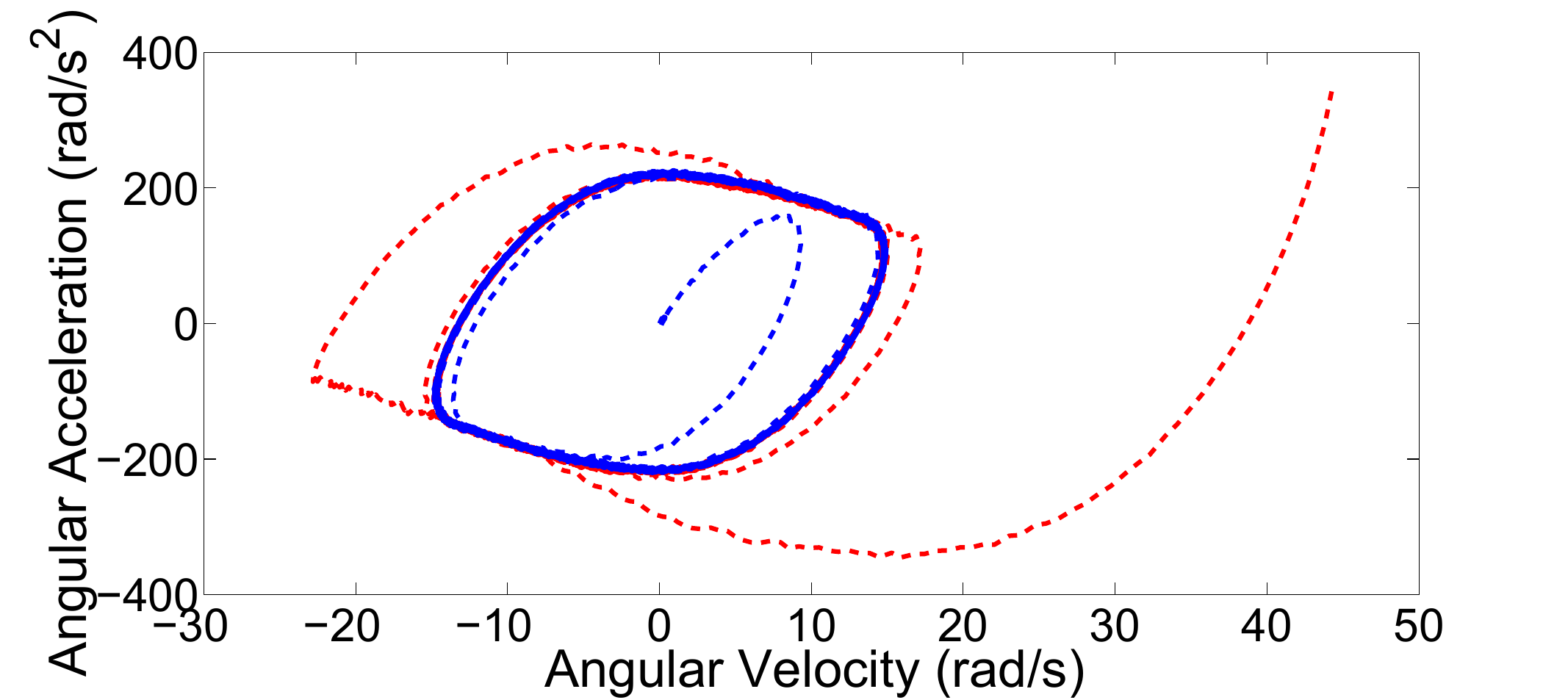}
\end{center}
\caption{Trajectories starting from various initial conditions converging to a limit cycle in phase space for delay $0.05$ $s$ and saturation limits $-1$ to $1$ $N$-$m$ (Servo motor experimental example). \label{limitcycle1}}
\end{figure} 

\begin{figure}
\centering
\includegraphics[scale=0.25]{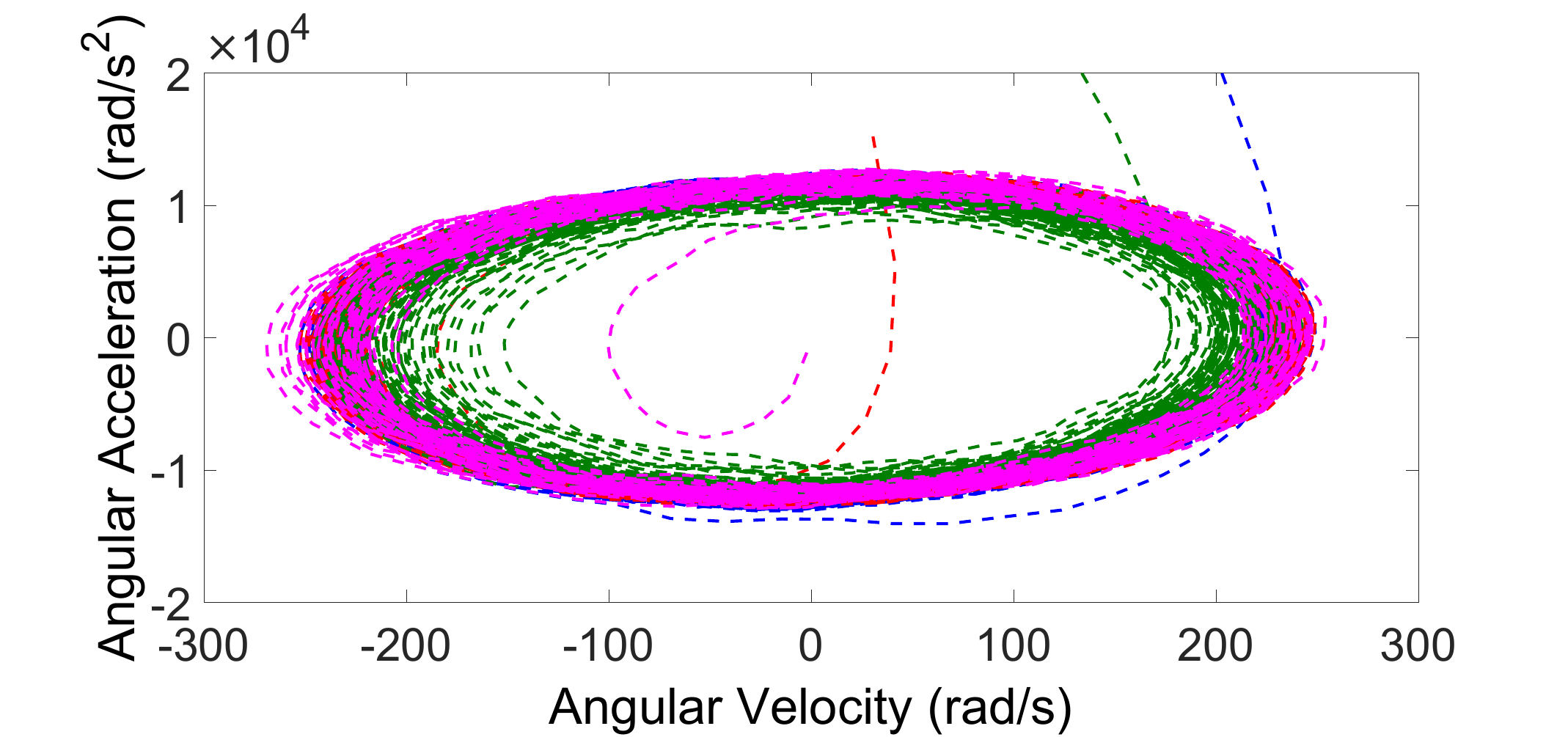}
\caption{Trajectories starting from various initial conditions converging to a limit cycle in phase space (Rotary inverted pendulum experimental example). \label{rotary}}
\end{figure}

\subsection{Rest tremor often disappears when large-scale voluntary motion is attempted \cite{34}.}
\indent Our numerical and experimental examples support this clinical observation by revealing that when a significant intended velocity is used (large-scale voluntary motion), the tremor disappears. Figures \ref{sim3} and \ref{ex3} illustrate this finding for the Pendulum numerical example and the Servo motor experimental example, respectively. As an example, the intended velocity is chosen to be a low-frequency sinusoidal input. In Figs. \ref{sim3} and  \ref{ex3}, the first subplots, which corresponds to zero input amplitude (implying no intended motion), reproduces the tremor. The traces of the (higher frequency) tremor is still present for a low amplitude input (second subplots) while the tremor visibly disappears for higher value of input amplitude (third subplots). This is further obvious in the frequency domain as shown in Fig. \ref{fft_sim3}. In the third subplot of Fig. \ref{fft_sim3}, in addition to the peak at the low frequency sinusoidal input, a new second peak is observed which is merely a harmonic of the low-frequency sinusoidal input and thus unrelated to tremor.\\
\indent This amplitude-dependent behavior is expected as a nonlinear effect of the saturation in the loop. The noteworthy finding is that all the three examples show similar trend of tremor disappearance with increasing amplitude of the intended velocity. This consistent trend needs a rigorous analysis as a nonlinear effect of the saturation.

\subsection{Rest tremor disappears when patients sleep or during mental concentration \cite{9}.}
When a patient is asleep or engaged in engrossing mental activity requiring mental concentration, one can argue that their sensory feedback is cut off or diminished. It is then obvious from the explanation in Sec. \ref{rest_tremor} that if the feedback path is disrupted, then the tremors should disappear. This explains why Parkinsonian rest tremor, in contrast to ``Essential tremor'' \cite[Table 2]{9}, disappears when patients sleep. One may argue that in the case in which a patient is undertaking an activity requiring mental concentration, the feedback may not be completely cut off but rather diminished. Since reducing the feedback gain (and consequently the loop gain) will have a stabilizing effect when there is a delay in the loop \cite{18,29}, this clinical fact is also explained. This argument further explains the surgical serendipity \cite{39, 40}, in which brain's control circuitry of a PD patient was accidentally damaged, and resulted in disappearance of rest tremor. This observation also gives a key insight that sensory feedback plays a crucial role in the mechanism causing rest tremors.
\\

\begin{figure}
\begin{center}
\includegraphics[scale=0.25]{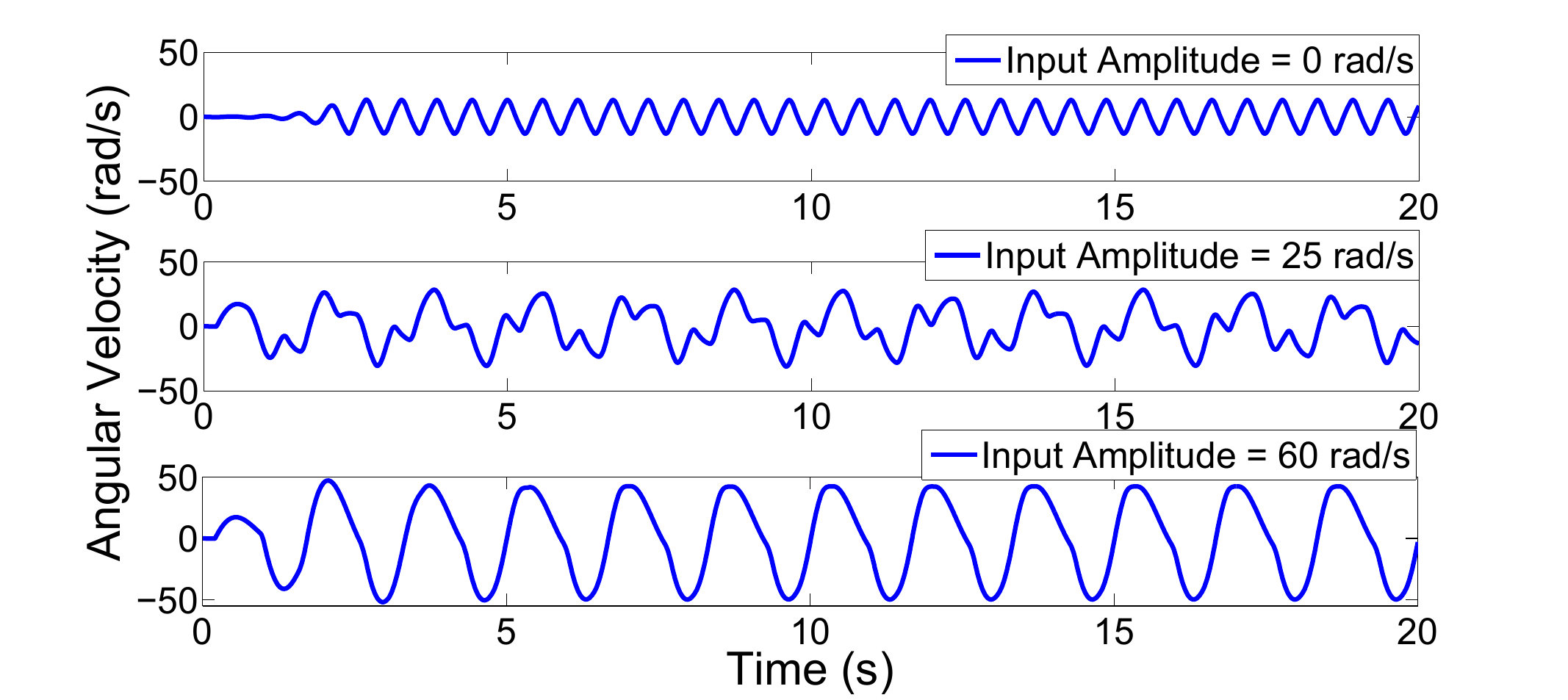}
\end{center}
\caption{Velocity plots for a sinusoidal intended velocity of frequency $0.6$ $Hz$, saturation limits $-100$ to $100$ $N$-$m$, delay $0.2$ $s$ and amplitude $0$, $25$ and $60$ $rad/s$ (Pendulum numerical example).\label{sim3}}
\end{figure}

\begin{figure}
\begin{center}
\includegraphics[scale=0.25]{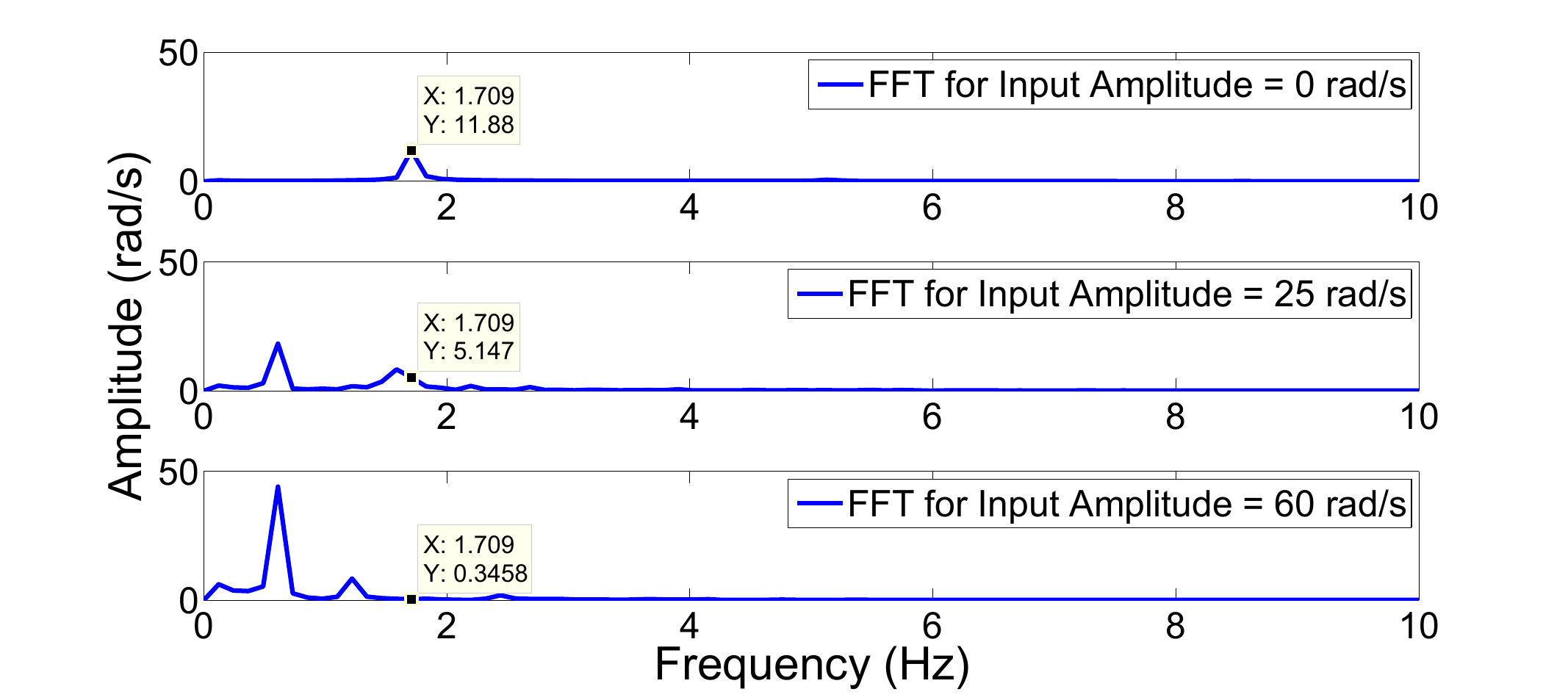}
\end{center}
\caption{Frequency content of velocity plots for a sinusoidal intended velocity of frequency $0.6$ $Hz$, saturation limits $-100$ to $100$ $N$-$m$, delay $0.2$ $s$ and amplitude $0$, $25$ and $60$ $rad/s$ (Pendulum numerical example). Here $x = 1.709$ $Hz$ is frequnecy of rest tremor and the amplitude of this rest tremor is subsequently reduced (second and third subplots) as we increased the input amplitude. Note that in third subplot, due to saturation nonlinearity, we observed the peak (at $x=1.2$ $Hz$) other than intended velocity frequnecy $0.6$ $Hz$ which is second harmonic of the intended velocity frequnecy $0.6$ $Hz$. \label{fft_sim3}}
\end{figure}

\begin{figure}
\begin{center}
\includegraphics[scale=0.25]{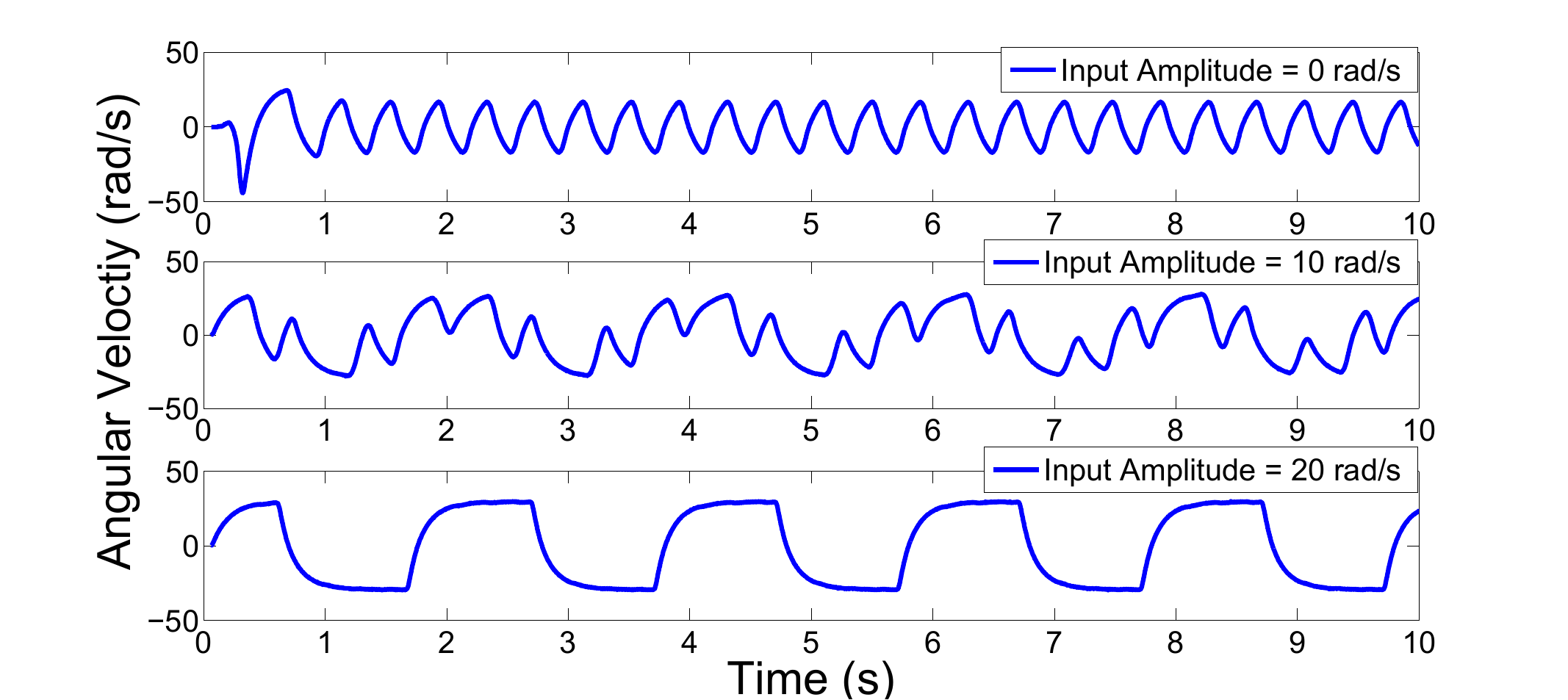}
\end{center}
\caption{Velocity plots for a sinusoidal intended velocity of frequency $0.5$ $Hz$, saturation limits $-1$ to $1$ $N$-$m$, delay $0.05$ $s$ and amplitude $0$, $10$ and $20$ $rad/s$ (Servo motor experimental example).\label{ex3}}
\end{figure}

\subsection{As the disease progresses, a decrease in frequency of tremor is observed and a corresponding increase in the amplitude of tremor is observed \cite{19}.}
 \indent From the explanation of the previous clinical facts, we conclude that the progression of the disease must be directly linked to increased response times. Further, all our numerical and experimental examples show a decrease in the frequency of oscillation (tremor) with an increased delay. Figures \ref{sim4} and \ref{ex4} show this finding for the Pendulum numerical example and the Servo motor experimental example, respectively. It is clear that there exist an inverse relationship between the loop delay and frequency of tremor and that further results in an increase in the amplitude as explained below. Consider a simple oscillating signal such as $a=sin(\omega t)$, where $\omega$ is an angular frequency and $a$ is an acceleration. Noting that its integral is $v=-cos(\omega t)/\omega $, it is obvious that for periodic oscillations the amplitude of the velocity signal is inversely proportional to the frequency of the acceleration signal. Here, since the amplitude of the acceleration signal is constrained by the saturation levels in control actions, it is not surprising that for the same saturation level, the amplitude of velocity signal is higher for lower frequency oscillations and vice versa. It should also be noted here that this explanation holds regardless of whether the plant has low-pass filter characteristics or not. 
\begin{figure}
\begin{center}
\includegraphics[scale=0.23]{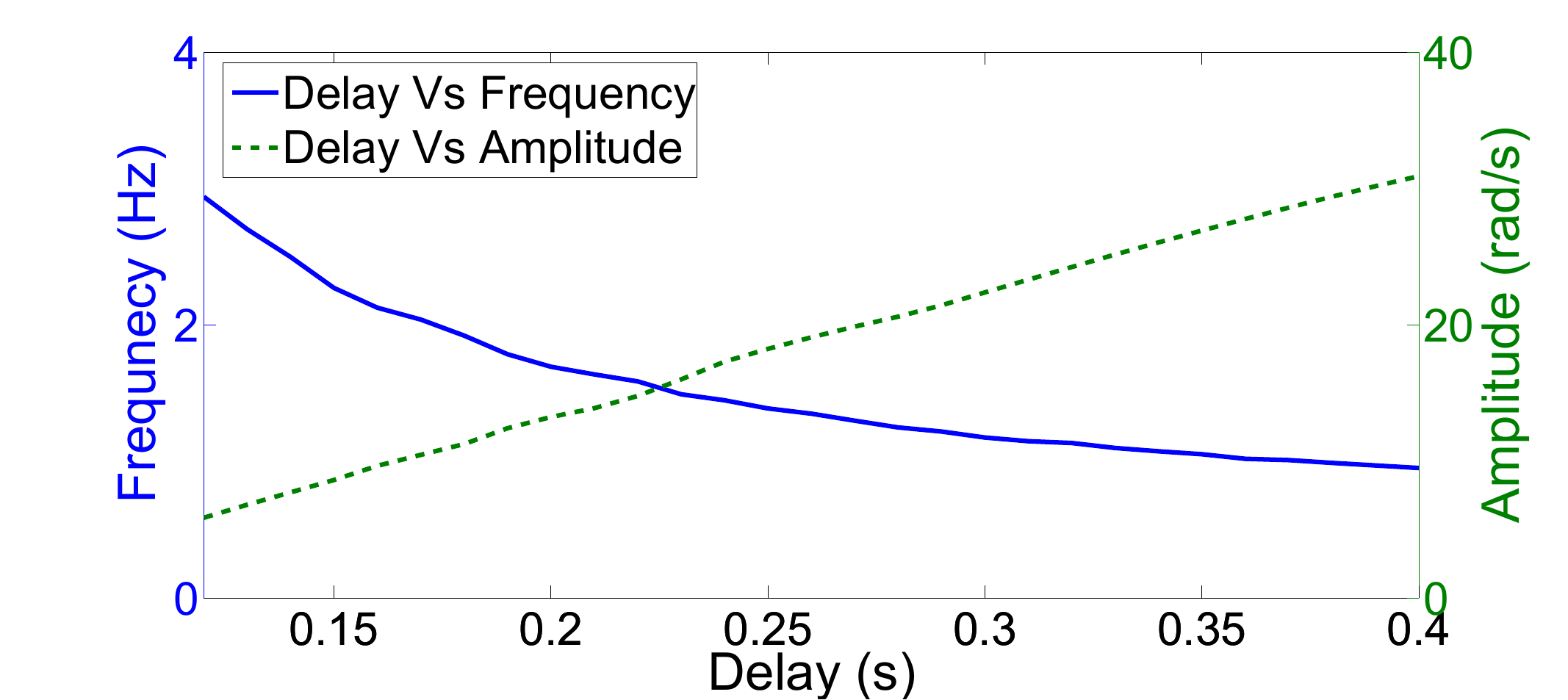}
\end{center}
\caption{Delay vs. frequency and amplitude of oscillation with saturation limits $-4$ to $4$ $N$-$m$ (Pendulum numerical example). \label{sim4}}
\end{figure}
\begin{figure}
\begin{center}
\includegraphics[scale=0.23]{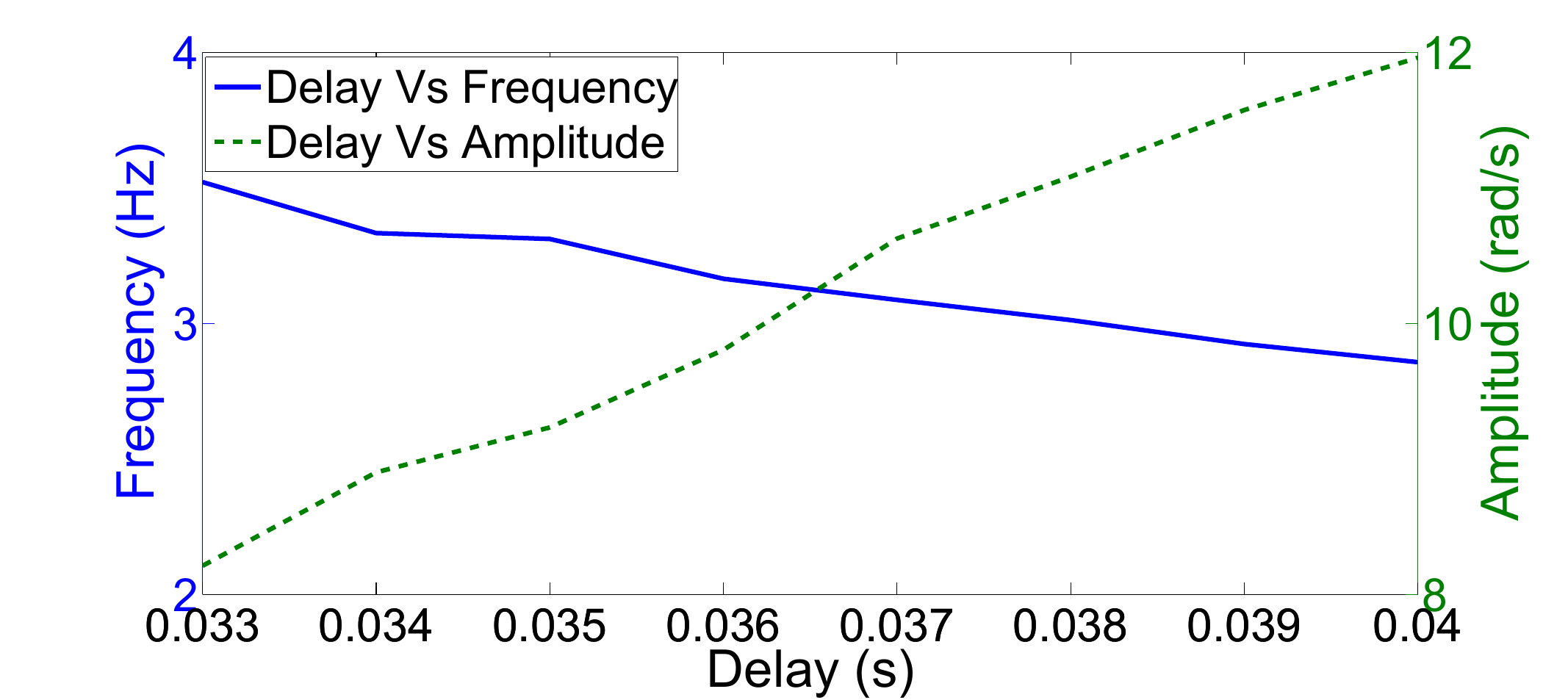}
\end{center}
\caption{Delay vs. frequency and amplitude of oscillation with saturation limits as $-10$ to $10$ $N$-$m$ (Servo motor experimental example). \label{ex4}}
\end{figure}

In the next section, we explore if these trends lead to specific features that can be used for diagnosis of Parkinson's disease and estimate its severity.

\section{Possibilities for Progress Tracking and Diagnosis}
\label{diagnosis}
From the discussion in the previous section, it is emerging clearly that there are trends in frequency and amplitude of tremors that can possibly be leveraged for tracking the progression of PD over a period of time and observing the effect of treatment strategies. For instance, one can develop a simple pocket device or a smartphone application (each described further in Secs. \ref{pocket_device} and \ref{smartphone_app} respectively), which use accelerometers and/or gyro sensors to record tremor data (accelerations and/or angular velocities) of PD patients. One could make measurements at regular intervals (say, every three months), and based on the frequency of the measured tremors, track the progression of the disease using trends such as observed in Figs. \ref{sim4} and \ref{ex4}. It may also be possible to evaluate the efficacy of treatment strategies using such an approach. 

In reality however, one would expect that it is very likely that PD tremors may show a combination of frequencies and may have some irregularities that could make it challenging to interpret the data. To resolve this issue, there can be two approaches. One approach is to explore other tremor parameters that may be more robust indicators of the underlying delay. A second approach is to take multiple independent measurements of different parameters that relate to the delay and then fuse the estimates (e.g. using a Maximum-likelihood method \cite{48}) to obtain a robust indication of the delay. This is a common practice in sensor fusion wherein several inexpensive low quality sensor measurements are fused to obtain a high-quality estimate.

On further exploration, two other parameters that correlate with the delay and therefore provide possibilities of tracking the progression of disease are the area and aspect ratio of the limit cycle (the closed trajectory formed in the phase space). The area is simply the area contained within the limit cycle, whereas the aspect ratio is defined as the ratio of angular acceleration range to angular velocity range. Since we know that for a harmonic signal, the aspect ratio is proportional to frequency, we may expect similar trends in aspect ratio and frequency. Figure \ref{5a} shows the trends in amplitude and frequency of the tremor, and area and aspect ratio of the limit cycle, as a function of delay and saturation for Pendulum numerical example. The frequency and aspect ratio are very sensitive to the delay, but are practically unaffected by the saturation while amplitude and area are affected by the saturation. Figure \ref{all_limits} demonstrate how the aspect ratio is directly linked to delay and relatively insensitive to saturation for all three examples. So, frequency and aspect ratio appear to be the most useful measurements among the four choices to track the delay. Nevertheless, it should be noted that amplitude, frequency, area of limit cycle and aspect ratio of limit cycle serves as four independent measurements that are correlated with the delay.

This is only a first step which will form the basis of patient studies that can help throw light on utility of these observations. The potential of using these parameters for progress tracking have to be confirmed through patient studies. A more challenging possibility that cannot be ruled out at the moment is to eventually use these parameters for diagnosis of PD. These would however have to be developed after extensive studies to factor out patient-to-patient variabilities.

\begin{figure} [h] 
\begin{center} 
\includegraphics[scale=0.211]{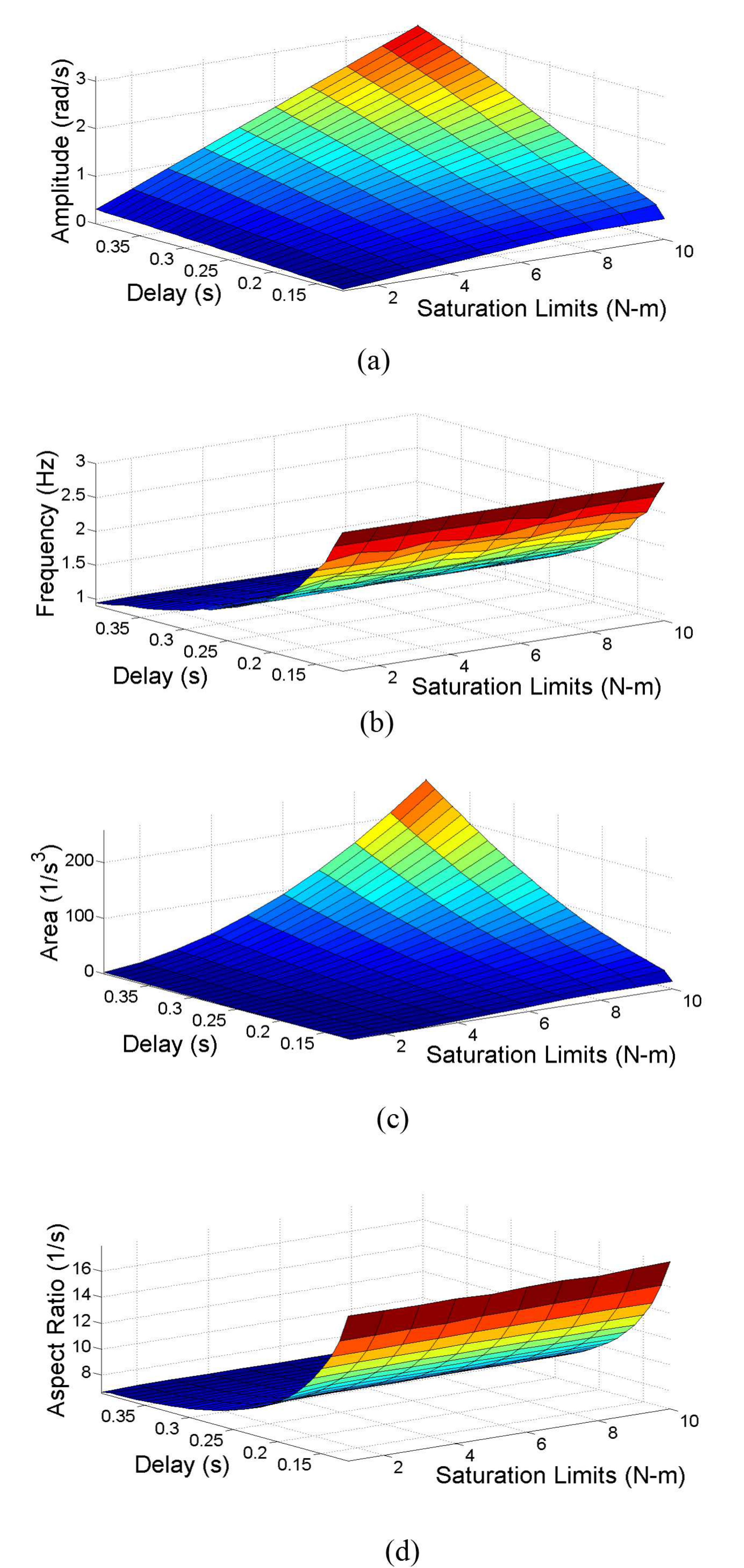}
\end{center}
\caption{(a) Amplitude of the tremor as a function of saturation and delay. (b) Frequnecy of the tremor as a function of saturation and delay. (c) Area of the limit cycle as a function of saturation and delay. (d) Aspect ratio of the limit cycle as a function of saturation and delay (Pendulum numerical example).\label{5a} }
\end{figure}

\begin{figure*}
\includegraphics[scale=1.074]{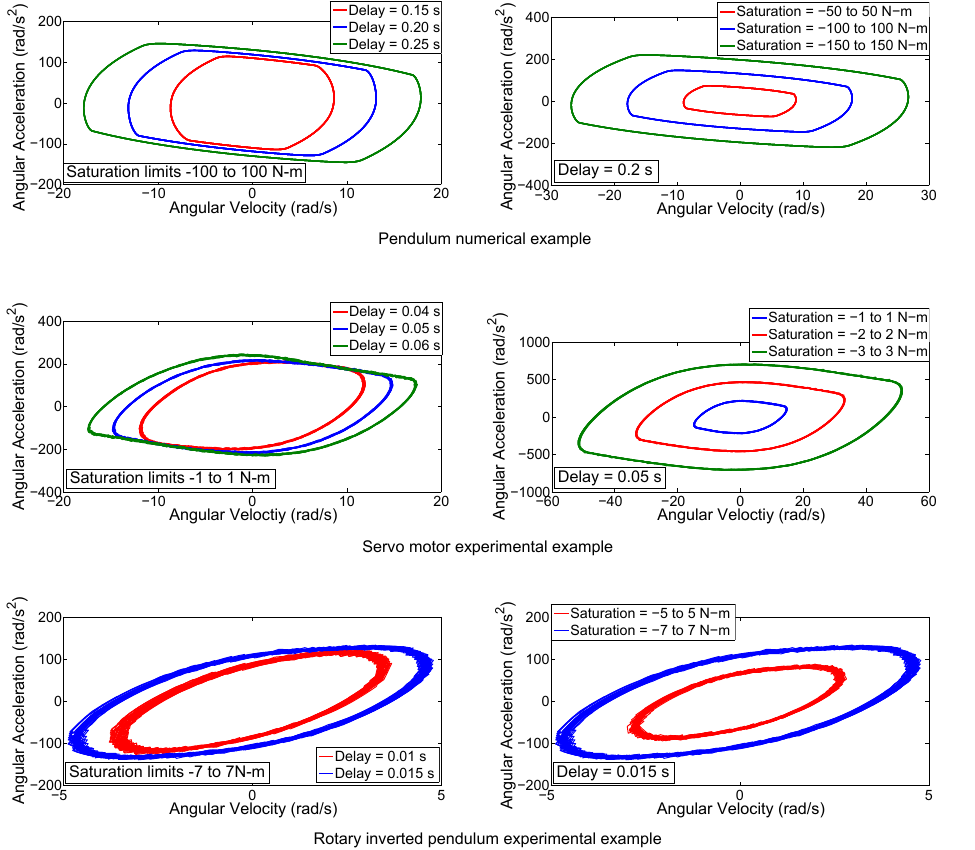}
\centering
\caption{ Effect of delay (left) and saturation (right) on the limit cycles. \label{all_limits}}
\end{figure*}

Nevertheless, in the following sub-sections we present two possible implementations based on the preliminary ideas discussed in Sec. \ref{diagnosis}, a pocket device, and a smartphone application. Both implementations utilize the angular velocity data of hand tremor from sensors (built-in sensors in the case of smartphone) and can potentially be used both for progress tracking and diagnosis. For diagnosis, a candidate algorithm is shown in Fig. \ref{flowchart} to detect presence and severity of PD.\\

\subsection{ Pocket Device \label{pocket_device}}

This pocket device consists of an accelerometer and a gyro sensor in a band that can be wrapped around the patient's hand (as shown in Fig. \ref{prototype}), along with a microcontroller and a display. The angular velocity of hand tremor measured by the gyro sensor is processed by the microcontroller as per the flowchart discussed in Fig. \ref{flowchart}. An advantage of such a device is its simplicity, cost and usability.
\begin{figure}
\begin{center}
\includegraphics[scale=.35]{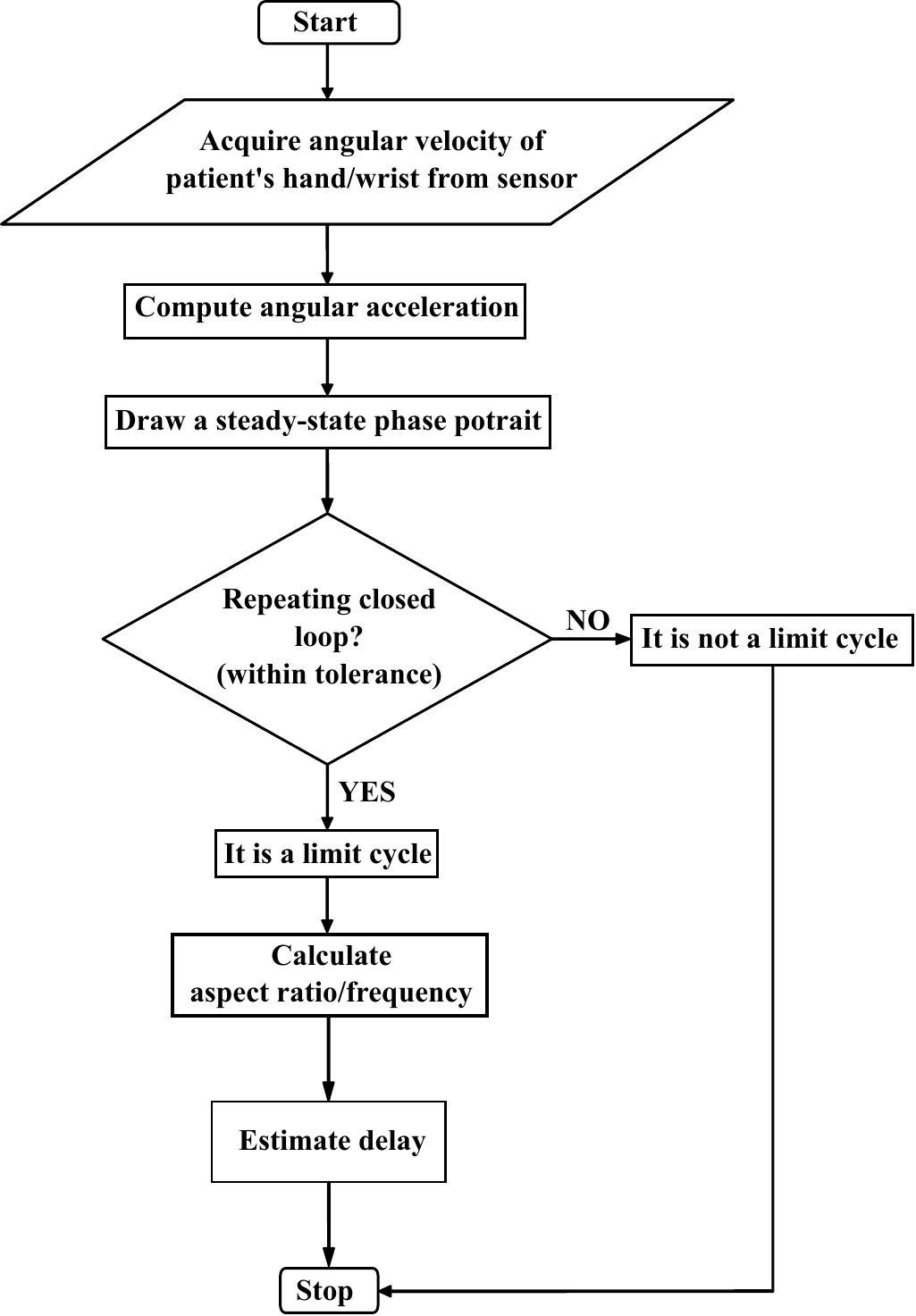}
\end{center}
\caption{A candidate algorithm for the diagnosis of Parkinson's disease.\label{flowchart}}
\end{figure}
\begin{figure}[h!]
\begin{center}
\includegraphics[scale=.25]{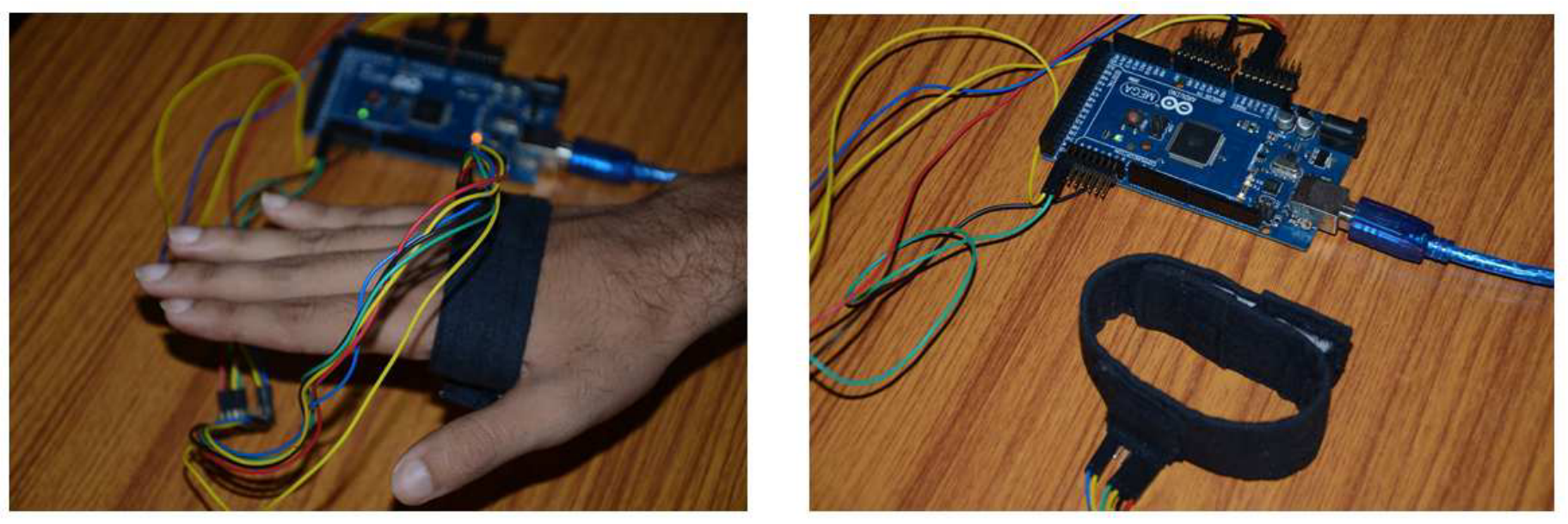}
\end{center}
\caption{A proof-of-concept prototype for a pocket device, which can further evolve into a convenient and compact wireless device.\label{prototype}}
\end{figure}\\\\
\subsection{Smartphone Application \label{smartphone_app}}

The same idea is also implemented on a smartphone application (Fig. \ref{mobileapp}). In this case, the patient can simply hold the smartphone in a predetermined configuration and the built-in gyro sensor of the smartphone is used to sense angular velocity of the hand/wrist, while the computations are performed on the smartphone processor and the results immediately displayed on screen. The user interface and additional features can be designed for a user-friendly and informative product, and can be used for telemedicine application.
\begin{figure} [h!]
\begin{center}
\includegraphics[scale=.4]{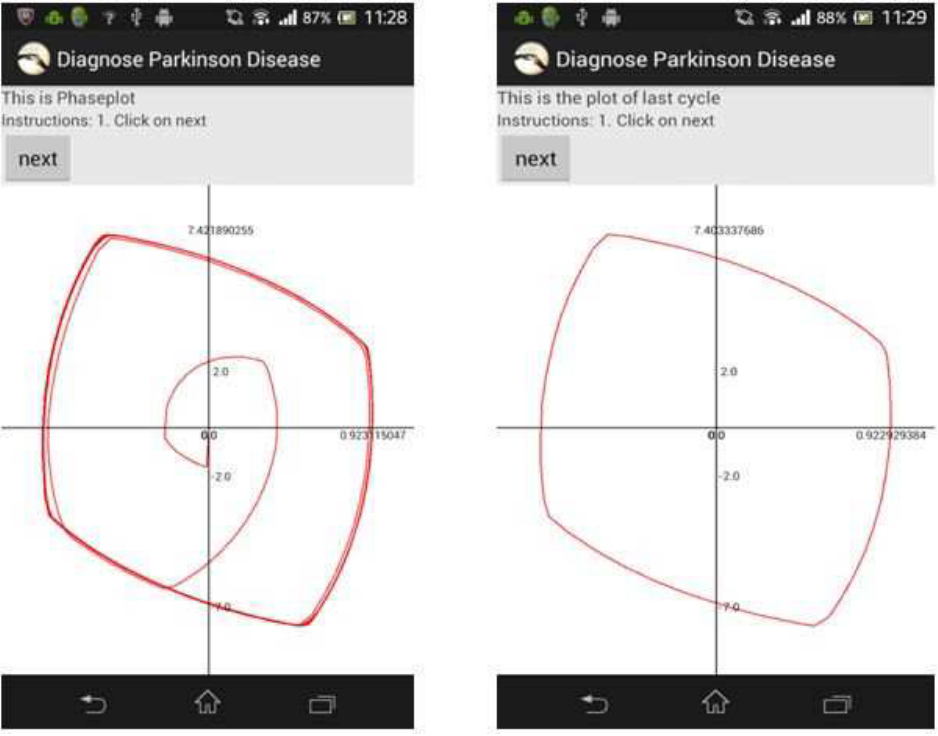}
\end{center}
\caption{Prototype for a smartphone application \cite{44}. \label{mobileapp}}
\end{figure}

\section{Conclusions}
\label{conclusion}
We started with the hypothesis that an increased response time is the cause of rest tremors in Parkinson's disease. With this starting point, and with the use of two simple experimental examples and a numerical example along with feedback control arguments, we were qualitatively able to explain several clinical observations related to rest tremors in PD. Thus, this work helps strengthen the hypothesis while laying a foundation for developing a better understanding of the mechanism behind rest tremors. Finally, using numerical and experimental examples, we further explored the possibility of using the tremor amplitude, frequency, area of limit cycle and aspect ratio of the limit cycle for tracking the progress of the disease and for diagnosis.

\begin{acknowledgment}
The authors gratefully acknowledge the support received from IIT Gandhinagar and the Ministry of Human Resources in the form of the fellowships for the first author.
The authors thank Taymaz Homayouni and Yousef Sardahi, who are Ph.D. students at University of California, Merced, for reviewing and editing this paper.
\end{acknowledgment}


\bibliographystyle{asmems4}

\bibliography{JCND}

\begin{thebibliography}{10}

\bibitem{11}
Palanthandalam-Madapusi, H.~J., and Goyal, S., 2011.
\newblock ``{Is Parkinsonian Tremor a Limit Cycle?}''.
\newblock {\em Journal of Mechanics in Medicine and Biology, {\bf 11}}(05),
  pp.~1017--1023.

\bibitem{9}
Jankovic, J., 2008.
\newblock ``{Parkinson's Disease: Clinical Features and Diagnosis}''.
\newblock {\em {Journal of Neurology, Neurosurgery \& Psychiatry}, {\bf
  79}}(4), pp.~368--376.

\bibitem{3}
Parkinson, J., 2002.
\newblock ``{An Essay on the Shaking Palsy}''.
\newblock {\em The Journal of Neuropsychiatry and Clinical Neurosciences}.

\bibitem{34}
Helmich, R.~C., Hallett, M., Deuschl, G., Toni, I., and Bloem, B.~R., 2012.
\newblock ``{Cerebral Causes and Consequences of Parkinsonian Resting Tremor: A
  Tale of Two Circuits?}''.
\newblock {\em Brain, {\bf 135}}(11), pp.~3206--3226.

\bibitem{35}
Hallett, M., 2012.
\newblock ``{Parkinson's Disease Tremor: Pathophysiology}''.
\newblock {\em Parkinsonism \& Related Disorders, {\bf 18}}, pp.~S85--S86.

\bibitem{49}
Santens, P., Boon, P., Van~Roost, D., and Caemaert, J., 2003.
\newblock ``{The Pathophysiology of Motors Symptoms in Parkinson's Disease}''.
\newblock {\em Acta Neurologica Belgica, {\bf 103}}(3), pp.~129--134.

\bibitem{50}
Marsden, C., 1989.
\newblock ``{Slowness of Movement in Parkinson's Disease}''.
\newblock {\em Movement Disorders, {\bf 4}}(S1), pp.~S26--S37.

\bibitem{2}
SAK, W., 1925.
\newblock ``{The Croonian Lectures on Some Disorders of Motility and of Muscle
  Tone, With Special Reference to the Corpus Striatum}''.
\newblock {\em The Lancet, {\bf 206}}(5314), pp.~1 -- 10.

\bibitem{13}
Heilman, K.~M., Bowers, D., Watson, R.~T., and Greer, M., 1976.
\newblock ``{Reaction Times in Parkinson Disease}''.
\newblock {\em Archives of Neurology, {\bf 33}}(2), pp.~139--140.

\bibitem{14}
Evarts, E., Ter{\"a}v{\"a}inen, H., and Calne, D., 1981.
\newblock ``{Reaction Time in Parkinson's Disease}''.
\newblock {\em Brain: A Journal of Neurology, {\bf 104}}(1), pp.~167--186.

\bibitem{15}
Paunikar, V.~M., Shastri, N., and H.Baig, M.~N., 2012.
\newblock ``{Effect of Parkinson's Disease on Audiovisual Reaction Time in
  Indian Population}''.
\newblock {\em International Journal of Biological and Medical Research, {\bf
  3}}(1), pp.~1392--1396.

\bibitem{17}
Bloxham, C., Dick, D., and Moore, M., 1987.
\newblock ``{Reaction Times and Attention in Parkinson's Disease}''.
\newblock {\em Journal of Neurology, Neurosurgery \& Psychiatry, {\bf 50}}(9),
  pp.~1178--1183.

\bibitem{6}
Beuter, A., 2003.
\newblock {\em Nonlinear Dynamics in Physiology and Medicine}.
\newblock Interdisciplinary Applied Mathematics. Springer.

\bibitem{37}
Moro, E., and Lang, A.~E., 2006.
\newblock ``{Criteria for Deep-Brain Stimulation in Parkinson’s Disease:
  Review and Analysis}''.
\newblock {\em Expert Review of Neurotherapeutics, {\bf 6}}(11),
  pp.~1695--1705.

\bibitem{38}
Modolo, J., and Beuter, A., 2009.
\newblock ``{Linking Brain Dynamics, Neural Mechanisms, and Deep Brain
  Stimulation in Parkinson's Disease: An Integrated Perspective}''.
\newblock {\em Medical Engineering \& Physics, {\bf 31}}(6), pp.~615--623.

\bibitem{10}
Austin, G., and Tsai, C., 1962.
\newblock ``{A Physiological Basis and Development of a Model for Parkinsonian
  Tremor}''.
\newblock {\em Stereotactic and Functional Neurosurgery, {\bf 22}}(3-5),
  pp.~248--258.

\bibitem{31}
Chagdes, J.~R., Rietdyk, S., Haddad, J.~M., Zelaznik, H.~N., Raman, A.,
  Denomme, L., and Cinelli, M.~E., 2013.
\newblock ``{Dynamic Stability of a Human Standing on a Balance Board}''.
\newblock {\em Journal of Biomechanics, {\bf 46}}(15), pp.~2593--2602.

\bibitem{45}
Chagdes, J.~R., Huber, J.~E., Saletta, M., Darling-White, M., Raman, A.,
  Rietdyk, S., Zelaznik, H.~N., and Haddad, J.~M., 2016.
\newblock ``{The Relationship Between Intermittent Limit Cycles and Postural
  Instability Associated With Parkinson's Disease }''.
\newblock {\em Journal of Sport and Health Science, {\bf 5}}(1), pp.~14 -- 24.

\bibitem{46}
Chagdes, J.~R., Rietdyk, S., Haddad, J.~M., Zelaznik, H.~N., Cinelli, M.~E.,
  Denomme, L.~T., Powers, K.~C., and Raman, A., 2016.
\newblock ``{Limit Cycle Oscillations in Standing Human Posture}''.
\newblock {\em Journal of Biomechanics}.

\bibitem{44}
Manasa, B., Dolores, J., Goyal, S., and Palanthandalam-Madapusi, H., 2014.
\newblock ``{Modeling and Simulations of Biomechanical Symptoms of Parkinson's
  Disease}''.
\newblock {\em Biophysical Journal, {\bf 106}}(2), pp.~794a--795a.

\bibitem{33}
Sun, J., and Voglewede, P.~A., 2014.
\newblock ``{Dynamic Simulation of Human Gait Using a Combination of Model
  Predictive and PID Control}''.
\newblock In ASME 2014 International Design Engineering Technical Conferences
  and Computers and Information in Engineering Conference, American Society of
  Mechanical Engineers, pp.~V006T10A008--V006T10A008.

\bibitem{47}
Sun, J., Wu, S., and Voglewede, P.~A., 2015.
\newblock ``{The Development of a Human Gait Model With Predictive Capability
  and the Simulation of Able-Bodied Gait}''.
\newblock In ASME 2015 International Design Engineering Technical Conferences
  and Computers and Information in Engineering Conference, American Society of
  Mechanical Engineers, pp.~V006T10A006--V006T10A006.

\bibitem{12}
Edwards, R., Beuter, A., and Glass, L., 1999.
\newblock ``{Parkinsonian Tremor and Simplification in Network Dynamics}''.
\newblock {\em Bulletin of Mathematical Biology, {\bf 61}}(1), pp.~157--177.

\bibitem{18}
Franklin, G., Powell, J., and Emami-Naeini, A., 2010.
\newblock {\em Feedback Control of Dynamic Systems}.
\newblock Alternative Etext Formats. Pearson.

\bibitem{29}
Beuter, A., B{\'e}lair, J., Labrie, C., and B{\'e}lair, J., 1993.
\newblock ``{Feedback and Delays in Neurological Diseases: A Modeling Study
  Using Dynamical Systems}''.
\newblock {\em Bulletin of Mathematical Biology, {\bf 55}}(3), pp.~525--541.

\bibitem{39}
Cooper, I.~S., 1953.
\newblock ``{Ligation of the Anterior Choroidal Artery for Involuntary
  Movements-Parkinsonism}''.
\newblock {\em The Psychiatric Quarterly, {\bf 27}}(1-4), pp.~317--319.

\bibitem{40}
Hornyak, M., Rovit, R.~L., Simon, A.~S., and Couldwell, W.~T., 2001.
\newblock ``{Irving S. Cooper and the Early Surgical Management of Movement
  Disorders}''.
\newblock {\em Neurosurgical Focus, {\bf 11}}(2), pp.~1--5.

\bibitem{19}
Hellwig, B., Mund, P., Schelter, B., Guschlbauer, B., Timmer, J., and
  L{\"u}cking, C., 2009.
\newblock ``{A Longitudinal Study of Tremor Frequencies in Parkinson's Disease
  and Essential Tremor}''.
\newblock {\em Clinical Neurophysiology, {\bf 120}}(2), pp.~431--435.

\bibitem{48}
Ljung, L., 1998.
\newblock {\em System Identification: Theory for the User}.
\newblock Springer.

\end{thebibliography}

 \appendix       
\section*{Appendix A: Analysis of location of sensorimotor loop delay}
\label{location}
Although we performed the analysis with the delay in the forward path (between the controller and plant) as indicated in Fig. \ref{Block_Diagram}, we note that the position of the delay in the loop doesn't affect the phase portrait and therefore our conclusions. To understand this, consider the case in which we have a delay ($t_{d2}$) in the feedback path that is additional to the delay ($t_{d1}$) in the forward path. In this case, under the linearity assumption, the output becomes, $ Y(s)=e^{-t_{d1}s}{\tilde{G}} R(S)$, where $\tilde{G}=\frac{G(s)} {1+G(s) e^{-t_ds}}$ is the closed loop transfer function with $t_d=t_{d1}+t_{d2}$. Note here that the nature of the response is determined by $\tilde{G}$ which has the total delay in its denominator, while $e^{-t_{d1}s}$ only serves to time-shift the output $Y(s)$. Therefore, in the phase space, the trajectories would only depend on the total combined delay in the loop and is unaffected by the actual positions of the delay elements in the sensorimotor loop. This claim needs to be further tested with simulations employing nonlinear plants and controllers. Thus, the knowledge of the mathematical model governing motor-control loop and details of the loop delay locations are not needed for diagnosis or for progress tracking.

\appendix       
\section*{Appendix B: Numerical and Experimental Examples}
\textbf{Numerical Example:} The plant that we use in the numerical example can be thought of as a simple pendulum with length $L$, mass $m$, and the damping coefficient $c$. We linearize the pendulum about its stable equilibrium (vertically downward) and write it in the state-space form as follows:

\begin{equation*}
\begin{split}
\dot{x} \hspace{2mm} = & \hspace{2mm} Ax + Bu,\\
 y \hspace{2mm} = &  \hspace{2mm} Cx + Du,\\
 \end{split}
\end{equation*}
where $x= [\begin{matrix}
\theta & \dot{\theta}
\end{matrix}]$ $\in R^{2}$ is the state vector with $\theta$ being the angle of the pendulum, $u \in R$ is the controlling torque on the pendulum as determined by the controller based on the feedback $y \in R$, which is the measured angular velocity of the pendulum, and 
\begin{eqnarray*}
A=\begin{bmatrix}
   0 & 1 \\
-g/L & -c/mL^2
\end{bmatrix},
B=\begin{bmatrix}
0 \\
1/mL^2
\end{bmatrix},
C=\begin{bmatrix}
0 & 1
\end{bmatrix}, D = 0.
\end{eqnarray*}

\indent For our simulation purpose, we use $L$ = $0.65$ $m$, $m$ = $3.5$ $kg$, $c$= $3.375$ $kg$-$m/s$, saturation limits of $-100$ to $100$ $N$-$m$ and a Proportional-Integral-Derivative (PID) controller with the proportional gain $k_P = 15$ $m$-$L^2$, integral gain $k_I = 4$ $m$-$L^2$ and derivative gain $k_D = 0.5$ $m$-$L^2$. The simulations are performed in Matlab SIMULINK.\\

\textbf{Experimental Examples:} We consider two motion control experimental examples, an angular position control experiment for a servo motor and a rotary inverted pendulum balancing experiment. The former is a first-order, linear, stable system with a proportional-derivative controller and the latter is a fourth-order, non-linear, unstable system with LQR controller. With these motion-control experiments, the controller parameters are tuned such that the closed-loop control system is stable analogous to the case of healthy individuals. Then we experimentally observe the effect of increasing delay by replacing the dynamics of simple pendulum (in case of simulation example) with the dynamics of servo motor or rotary inverted pendulum as body dynamics (Fig. \ref{Block_Diagram}). Both of these experiments are based on a QUBE rotary servo experiment from QUANSER as shown in Fig. \ref{fig:quarc}.\\
\renewcommand{\thefigure}{B.\arabic{figure}}
\setcounter{figure}{0}    
\begin{figure}
\begin{center}
\includegraphics[scale=0.42]{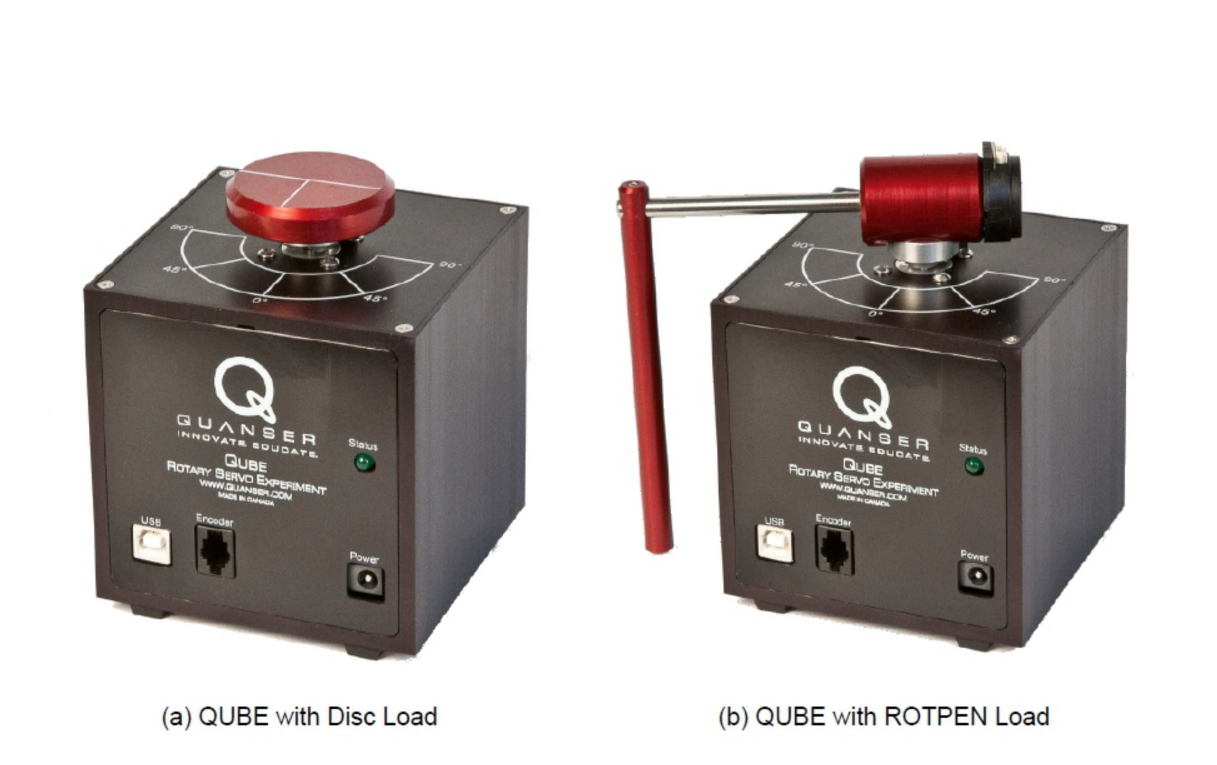}
\end{center}
\caption{QUBE rotary servo experiment setup. (a) Position control experiment setup (left) and (b) Inverted pendulum setup (right).\label{fig:quarc}}
\end{figure}
\indent It is worthwhile to note here that since the focus of the paper is on qualitative observations relating to Parkinsonian tremor, the values of the parameters in the examples are indicative values and not related to parameters of a real human body.

\end{document}